\def\be{\begin{equation}}
\def\ee{\end{equation}}
\def\bea{\begin{eqnarray}}
\def\eea{\end{eqnarray}}
\def\source#1#2#3#4{{\it #1}~{\bf #2}, #3 (#4)}
\def\Eq#1{Eq. \ref{#1}}

\def\ie{{\it ie}}

\def\irt{\frac{1}{\sqrt{2}}}

\def\ket#1{| #1 \rangle}

\def\Psipm{|\Psi_{\pm}\rangle} 
\def\Psimp{|\Psi_{\mp}\rangle} 
\def\PsipmPhi{|\Psi_{\pm}(\Phi)\rangle} 
 
\def\Psipmnk{|\Psi_{\pm}\setnk \rangle} 
\def\Psipmarg#1{|\Psi_{\pm}(#1)\rangle}

\def\Psitpm{|\tilde{\Psi}_{\pm} \rangle}
\def\Psitp{|\tilde{\Psi}_+ \rangle}

\def\Psitpnk{|\tilde{\Psi}_+\setnk \rangle}

\def\zk{z_k}
\def\Zk{Z_k}
\def\Xk{X_k}
\def\Yk{Y_k}
\def\nk{n_k}
\def\setnk{\{n_k\}}
\def\setbnk{\{\bar{n}_k\}}
\def\phik{\phi_k}
\def\Rphik{R(\{\phik\})}
\def\Rinvphik{R^{-1}(\{\phik\})}
\def\CO{{\cal O}}
\def\COphik{{\cal O}(\{\phik\})}
\def\COphikp{{\cal O}(\{\phik'\})}
\def\CN{{\cal C}_N}
\documentstyle[aps, prl, graphics, twocolumn, epsfig, floats]{revtex}
\title{Rotational Properties and GHZ Contradictions in N-Qubit Systems}
\author{Jay Lawrence \\ {\it Department of Physics, Dartmouth
College, Hanover, NH 03755}}

\date{\today}

\begin{document}
\maketitle
\begin{abstract}
Rotational symmetries of $N$-qubit Greenberger-Horne-Zeilinger (GHZ)
states directly exhibit their nonlocality and render transparent the many 
possible measurements that produce absolute contradictions with local 
realism.  While $N$ measurements fix the assumed elements of reality, an 
exponentially growing number $\CN = 2^{N-2} - 2^{(N-2)/2}\sin(N\pi/4)$ 
of absolute contradictions occurs.     
Operators that represent the detector settings provide a faithful 
representation of the symmetry group, and demonstrate Kochen-Specker 
contextuality.   
\end{abstract}
\bigskip
\centerline{PACS numbers:  03.65.Ud, 03.65.Ta, 03.67-a}
\bigskip
A measurement on one member of an entangled Bell pair determines the 
state of the other.  It is natural to ask whether a {\it rotation} 
of one {\it changes} the state of the other.  In fact the rotation 
changes the state of neither particle - it only changes the state of 
{\it both} particles!  That is, a measurement on either particle by 
itself registers no change (the outcomes remain random), while the 
perfect correlation between these outcomes changes in a predictable way. 
This reflects a symmetry of the Bell states:   Every transformation 
induced by rotating one qubit can be duplicated (or reversed) by 
rotating the other.  More concisely, one-qubit rotations are 
untraceable!    A corresponding symmetry exists within a subgroup 
$SO(2)^{\otimes N}$ of one-qubit rotations in $N$-particle 
Greenberger-Horne-Zeilinger (GHZ) states:  Each qubit, though 
completely unpolarized, has a special axis about which rotations 
are untraceable.   This symmetry leads directly, 
for all $N \geq 3$, to GHZ contradictions \cite{GHZ,Mermin1}, 
referring to contradictions between quantum mechanics and two 
distinct assumptions about hidden variable theories:  (a) {\it local 
realism} - that the predictability of a measurement outcome on an 
isolated object implies a pre-existing value (an ``element of reality'' 
in the sense of Einstein, Podolsky and Rosen \cite{EPR}), and (b) 
{\it noncontextuality} - that several observables, not all of which 
commute, are capable of simultaneously taking definite values, even 
though some of these must be statistically distributed.   
Contradictions of type (a) may be called ``Bell-EPR'' theorems, 
and of type (b), Kochen-Specker (KS) theorems \cite{KS}.

It is interesting to note that the original GHZ argument \cite{GHZ} 
was based upon the rotational properties of a four-particle (GHZ)
state.  It was shown that these properties cannot be reproduced by
local hidden variables, and (in its crucial step beyond the original 
Bell theorem \cite{Bell}) that particular rotations allow 
nonprobabilistic quantum statements, thus conforming to the 
definition of ``elements of reality'' \cite{EPR}, while also 
contradicting local realism.  Mermin reformulated the argument 
\cite{Mermin1} in terms of operator identities for three spin-1/2 
particles, from which a KS as well the Bell-EPR theorem immediately 
followed.  He subsequently derived violations of Bell 
inequalities \cite{Mermin2} for $N$ spin-1/2 particles, and much 
has developed along these lines since.   For example, Bell 
inequalities have been linked more deeply to local realism 
\cite{WWZukB}, to multiparticle entanglement \cite{Seevinck}, and
more practically to communication complexity \cite{B3} 
and the security of multi-party key distributions \cite{Sen}.  
Further KS operator identities have also been derived 
\cite{Pagonis,Cerf,PDiV}, and  related intimately to 
quantum error correction \cite{PDiV}.  

Experimental groups have reported the preparation and analysis of 
GHZ states with three photons \cite{BouwPan}, with two 
atoms and a cavity photon \cite{Rausch}, and with four ions 
\cite{Sackett}.  Recently, GHZ contradictions were observed in a 
four photon experiment \cite{Zhao} showing sufficient violations 
of Bell inequalities to establish four-particle entanglement 
unambiguously \cite{Seevinck}.

In this work, we trace GHZ contradictions to a symmetry of 
$N$-qubit GHZ states - specifically, the untraceability of 
uniaxial one-qubit rotations.  Despite its use in the original  
argument \cite{GHZ}, rotational symmetry has not been exploited 
to derive either the state-dependent contradictions or the 
corresponding operator identities for all $N$.   While general 
rotational invariance of the system has been used to constrain 
local realistic models \cite{Nagata}, the state dependent symmetry 
discussed here is useful in revealing specific measurements where 
GHZ contradictions are to be found.  And, hopefully,
analogous symmetries will prove useful in systems of many 
qu{\it dits} ($d$-state particles), where examples of Bell 
inequalities \cite{FuKasZuk} and KS operator identities 
\cite{Cerf} have been found.

Let us focus first on the state-specific Bell-EPR case 
because of its relevance to experimental realizations. 
To this end, consider the $N$-qubit GHZ states,
\be
  \Psipm = \irt \big( \ket{00...0} \pm \ket{11...1} \big),
\label{GHZ1}
\ee
where indices 0 and 1 denote spins ``up'' and ``down'' along the 
$\zk$-axes ($k =$ 1,...,$N$), whose orientations are arbitrary.  
Imagine an operation in which each qubit is rotated through an 
angle $\phik$ about its own $\zk$-axis, as effected by the operator 
\be
  \Rphik = \prod_{k} \exp (-i Z_k \phik /2),
\label{Rot}
\ee
where $\Zk$ is the Pauli matrix ($\sigma_z^k$) acting on the 
$k$-th qubit.  The rotated states $\PsipmPhi$ depend on the $\phik$'s
only through their sum $\Phi = \sum_{k} \phik$:
\bea
  \PsipmPhi & \equiv & \Rphik \Psipm    \label{rotpsi1}  \\
   & = & \big( \cos \Phi/2 \Psipm - 
             i \sin \Phi/2 \Psimp \big),
\label{rotpsi2}
\eea
so that rotating a single qubit transforms the state of the 
{\it system} without changing the state of the qubit itself, which 
remains random.  The encoded information ($\Phi$) can be extracted 
only by measuring all $N$ qubits, and individual rotations (about 
$z_k$) are untraceable.

The transformation induced by rotating one qubit may be undone by 
rotating another to make $\Phi = 0$.  Zurek \cite{Zurek} has used a
similar invariance to derive Born's probability rule for a system 
${\cal S}$ entangled with its environment ${\cal E}$.    

Note also that rotating all of the particles simultaneously through 
the same angle (say $\Phi/N$) has the same effect as rotating any 
one of them through $\Phi$.  This ``rotational wavelength'' 
compression is analogous to the translational wavelength compression 
recently observed in multi-photon states \cite{Shih,Walther}.   

The rotations of \Eq{Rot} form a group, $SO(2)^{\otimes N}$, and the 
states transform simply according to a two-dimensional representation 
(\Eq{rotpsi2}).  So the state transformations are isomorphic to those
of a single qubit prepared in the ``up'' or ``down'' state along the 
$x$-axis (\ie, $\Psipm$ with $N=1$), and then rotated about the $z$ 
axis.  For example, the states $\ket{\Psi_{\pm}(\pi/2)}$ correspond to 
``up'' and ``down'' states along  $y$, and a full $2\pi$ ``rotation'' 
produces the expected ($-$) sign for all $N$.  If the qubit is photon
polarization, then rotations are on the Bloch sphere: $\Psipm$ are 
$\pm 45^o$ linear polarization states, and $\ket{\Psi_{\pm}(\pi/2)}$ 
are right and left circular.  

Consider an operator of which $\Psipm$ are eigenstates,
\be
  \CO = \prod_k X_k,
\label{prodX}
\ee
where $X_k$ is the Pauli matrix ($\sigma_x^k$) for the $k$-th qubit.  
Since every $X_k$ factor interchanges 0 and 1 states of the $k$-th 
qubit with phase factors 1, the eigenvalues of $\Psipm$ are $\pm1$,
respectively.  We define rotations of this (and other operators) that 
preserve their expectation values in the rotated states, as
{\it co-rotations}:
\bea
   \COphik & \equiv & \Rphik \CO \Rinvphik     \label{rotop1}  \\
    & = & \prod_k(X_k \cos \phik + Y_k \sin \phik),
\label{rotop2}
\eea
where $Y_k$ denotes $\sigma_y^k$.   Since the rotated GHZ states 
$\PsipmPhi$ depend only on the collective angle $\Phi$, there are 
many co-rotations $\COphik$ of which they are eigenstates:  For 
{\it any} set of one-qubit rotations {$\phi_k'$} that sum to 
$\Phi$, Eqs. \ref{rotpsi1} and \ref{rotop1} show trivially that
\be
  \COphikp \PsipmPhi = \pm  \PsipmPhi, \hskip.5truecm
  \bigg( \sum \phi_k' = \Phi \bigg).
\label{rotop3}
\ee
Examples are shown in Fig. 1, where all points on the circles 
represent rotated (+) states $\ket{\Psi_+(\Phi)}$, and, at the 
special points $\Phi = 0$, $\pi/2$, $\pi$ and $3\pi/2$, we list the 
simplest ``co-rotated'' operators of which $\ket{\Psi_+(\Phi)}$ is 
an eigenstate with eigenvalue (+1).  States diametrically opposed to
an operator, $\ket{\Psi_+(\Phi + \pi)} \equiv -i \ket{\Psi_-(\Phi)}$, 
are eigenstates with eigenvalue ($-1$).  The two states $\Psipm$ 
correspond to $\Phi = 0$ and $\pi$ (E and W poles, respectively), 
while the two states at $\Phi = \pi/2$ and $3\pi/2$ (N and S poles) 
are
\begin{figure}
\centerline{\psfig{width=8cm,file=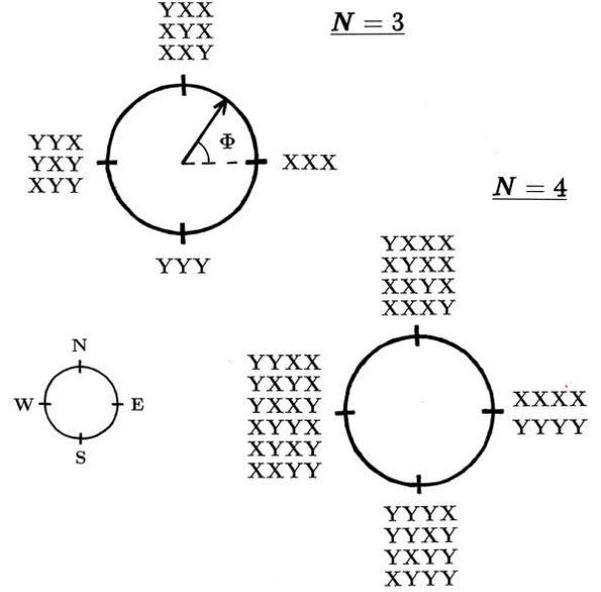}}
\caption{Rotated GHZ states are eigenstates of operators at the
same pole (\ie, co-rotated operators) with eigenvalue +1, and of 
operators at the opposite pole with eigenvalue $-1$.} 
\label{figure1}
\end{figure}
\be
  \Psitpm  \equiv  \Psipmarg{\pi/2} = 
  {1 - i \over 2} 
  \bigg( \ket{00...0} \pm i \ket{11...1} \bigg).            
\label{GHZ2}
\ee
So, for example, operators at N have $\Psitpm$ as eigenstates with 
eigenvalues $\pm1$, while operators at S have the same eigenstates 
but with opposite eigenvalues.

As a physical interpretation, each operator in Fig. 1 represents a 
distinct array of one-qubit detector settings, several of which could 
be used to analyse a particular state.  In the state $\Psitp$, for 
example, while all individual spin components are random, those 
products indicated by operators at N are always (+1), and at S, ($-1$).   
Since the corresponding detector arrays all differ by rotations of 
individual detectors (physical rotations for spin-1/2 particles, or 
Bloch sphere rotations for photons, involving quarter wave plates), 
one could, equivalently, fix the detector array and rotate individual 
qubits.  All operators at N are then related by compound rotations 
under which $\Psitp$ is invariant (\ie, $\Delta \Phi = 0$), while 
operators at N and S are related by compound rotations with 
$\Delta \Phi = \pi$, which produce a sign change in the product of 
spin components defined by the fixed detector array.  Local hidden 
variables cannot reproduce this behavior:  Constrained by the 
invariance of measurement outcomes under $\Phi$-preserving compound 
rotations, they predict invariance under {\it all} local 
rotations about $z_k$ axes.   

Specific GHZ contradictions reside in the eigenvalues of polar 
opposites (operators at N-S, {\it or} at E-W).   These eigenvalues are 
inconsistent with the assumption that the individual spin components 
take definite values (existing as ``elements of reality'' if the 
particles are separated spatially).   To review the familiar case of
$N=3$, consider the state $\Psitp$:   All three operators at the N pole 
take the value (+1).  Every choice of values for the individual factors 
($\pm 1$ for the $X_k$ and $Y_k$) consistent with this predicts (+1) 
for the product at S, the exact opposite of the quantum prediction.    

In the analogous state with $N=4$, all four operators at N  
take the value (+1), and now the assumption of local realism predicts 
the same value (+1) for {\it all four} operators at S, a fourfold 
contradiction with quantum theory.   With larger $N$ (as we shall show), 
knowing the values of only a {\it subset} of operators at the N pole 
(those, for example, having just a single factor of $Y$), local realism 
predicts values of {\it all} operators at S, every one in contradiction 
with quantum theory.   The number of contradictions is therefore 
equal to the number of operators at S:  Since S operators contain 3, 7, 
11,... factors of $Y$, the number of each type being given by a 
binomial coefficient, their total number is simply
\bea
    \CN = \left( {N \atop 3} \right) & + &
    \left( {N \atop 7} \right)  + ...  \nonumber     \\
     & = & 2^{N-2} - \sqrt{2^{N-2}}\sin(N\pi/4).
\label{viol}
\eea 
Table 1 lists $\CN$ {\it vs} 
$N$ to show how it oscillates about the trend $2^{N-2}$, which is about 
one-fourth of the total number ($2^N - 1$) of compatible observables.
The formula correctly yields ${\cal C}_2 = 0$, as Fig. 1 would suggest. 

The exponential growth of $\CN$ requires the entry of
``irreducible'' contradictions \cite{Pagonis,Cerf} (those which could 
not arise from fewer than all $N$ particles), which occur at 
$N=3,~7,~11,...$ .  So in  the case of $N=4$, the four contradictions 
arise from the possible groupings of irreducible 3-particle 
contradictions, as is evident in Fig. 1.  The entanglement itself,
of course, involves all 4 particles irreducibly:  The measurement 
outcomes on individual particles are random, with only the last being
predictable from the first three.  Such correlations were indeed 
observed \cite{Zhao} in connection with the experimental 
demonstration of genuine 4-particle entanglement.
 
\begin{table}
$$
\vbox{
  \halign{
  # \hfil & \hfil # \hfil & \hfil # \hfil & \hfil # \hfil & \hfil  
    # \hfil & \hfil # \hfil & \hfil # \hfil & \hfil # \hfil & \hfil   
    # \hfil & \hfil # \hfil                                         \cr
\noalign{\hrule}
\noalign{\smallskip}
  number of qubits  &  3 & 4 & 5 & 6 & 7 & 8 & 9 & 10 \cr
\noalign{\smallskip}
\noalign{\hrule}
\noalign{\smallskip}
  contradictions ${\cal C}_N$  & 1 & 4 & 10 & 20 & 36 & 64 & 120 & 240 \cr
  compatible obs. & 7 & 15 & 31 & 63 & 127 & 255 & 511 & 1023          \cr 
\noalign{\smallskip}
\noalign{\hrule}
}}
$$
\caption{The number $\CN$ of GHZ contradictions arising from 
operators at opposite poles in plots like Fig. 1.}
\end{table}
The above contradictions are derived, ostensibly, from a particular 
state, but {\it any} GHZ state can be mapped into this state by 
rotating individual qubit axes.  It follows that every GHZ state 
elicits $\CN$ contradictions. 

As a useful illustration of this point, we now show that Fig. 1 
accommodates complete bases of GHZ states and leads immediately to 
operator identities.  To begin, consider the basis that includes
\Eq{GHZ1} states,
\be
  \Psipmnk = \irt \bigg( \ket{\setnk} \pm \ket{\setbnk} \bigg),
\label{GHZ3}
\ee
where $\setnk$ is an $N$ digit binary number, and $\setbnk$ is its
complement \cite{binaries}.   Since any of these can be mapped 
into $\Psipm$ (Eq. \ref{GHZ1}) by flipping $z_k$-axes 
for which $n_k = 1$ \cite{flip}, the rotation identity 
(\Eq{rotpsi2}) applies provided that the sense of the corresponding
rotation angles $\phik$ is reversed in defining the collective angle,
\be 
 \Phi = \sum_k (-1)^{n_k} \phik.
\label{genrot}
\ee
Thus every basis state transforms according to a unique pattern of 
one-qubit rotations (and a unique 2D representation of 
$SO(2)^{\otimes N}$), which in 
turn defines the appropriate co-rotations of the operator $\CO$.  
Noting that {\it every} basis state (\Eq{GHZ3}) is an eigenstate of 
$\CO$  with eigenvalue $\pm1$ (recall every $X_k$ factor interchanges 
$0_k$ and $1_k$ with phase factors 1), the co-rotated operator,  
\be
  \CO\big( (-1)^{n_k}\phik \big) = \prod_k\big(X_k \cos \phik +
                        Y_k (-1)^{n_k} \sin \phik\big),
\label{rotop4}
\ee  
has eigenstates $\ket{\Psi_+(\setnk,\Phi)}$ with eigenvalues 
($+1$).  These rotations generate exactly the same operators shown 
in Fig. 1, but with sign reversals that indicate eigenvalue reversals.  
Thus, Fig. 1 applies to the state $\ket{\Psi_+\setnk(\Phi)}$ with the 
understanding that an operator's eigenvalue is reversed if it contains 
an odd number of $Y_k$ factors for which $n_k = 1$.

It follows that operators at the N pole carrying a single $Y$-factor 
(henceforth called $\CO_k$), have eigenvalues that uniquely label the 
{\it states} at N:
\be
   \CO_k \Psitpnk = (-1)^{n_k} \Psitpnk,
\label{index}
\ee
where these ``$\pi/2$'' states are (compare \Eq{GHZ2}),
\be
   \Psitpnk = {1 - i \over 2} 
  \bigg( \ket{\setnk} + i \ket{\setbnk} \bigg).  
\label{GHZ4}
\ee
Here, the $2^N$ binary numbers $\setnk$ label an (alternative) 
complete orthonormal set of GHZ states, all residing at N  
\cite{caveat}.   For each of these states, the eigenvalue of any
operator at N or S is given by the following simple rule:  Start 
with $+1$.  Multiply by ($-1$) for every $Y_k$ factor that has 
$\nk=1$.  Multiply again by ($-1$) for all S operators (because 
$\Psitpnk$ is an N state).  Thus, the eigenvalue of every such 
operator, $\CO_{kl...p}$ (where subscripts identify the $Y$-factors), 
is a product of eigenvalues of appropriate $\CO_k$'s with an 
additional ($-$) for S operators.  Since this rule holds for every 
basis state, it reflects an operator identity:  Every operator at N 
or S, each of which has an odd number ($2n+1$) of $Y$ factors, 
obeys
\be
  \CO_{kl...p} = (-1)^n \CO_k\CO_l...\CO_p,
\label{multrule}
\ee 
where $(-1)^n$  is negative for S operators because odd $n$ 
corresponds to 3, 7, ... $Y$ factors. 

The same observables $\CO_k$ serve as the vehicle for the assumptions 
of noncontextuality and local realism.   The operative assumption is 
that the spin components $X_k$ and $Y_k$ take definite values 
($\pm 1$), independently of what is being measured in a particular 
experiment.  So tensor product operators, such as $\CO$, take values 
$v(\CO)$ dictated by those of their factors,
\be
     v(\CO) = v(X_1)v(X_2)...v(X_N).
\label{value1}
\ee
The operators $\CO_k$ with single $Y$ factors take values  
\be
  v(\CO_k) = v(Y_k)v(X_k)v(\CO),
\label{value2}
\ee
since $v^2(X_k) = 1$.  Operators with any odd number of $Y$ factors 
(those at N and S poles) take values
\be
  v(\CO_{kl...q}) = v(\CO_k)v(\CO_l)...v(\CO_q),
\label{value3}
\ee
since all such products generate the common background factor 
$v(\CO)$.  Equations \ref{value3} and \ref{multrule} demonstrate 
sign contradictions for all operators at S, of which there are $\CN$. 

To show that E-W operators produce the same number of contradictions,
note that E-W operators are obtained from N-S operators by 
interchanging $\Xk$ and $\Yk$ in any {\it odd number} of qubits (for
odd $N$ it could be all of them).  Since this transformation is 
unitary ($\pi$ rotation about the 45$^o$ axis), the multiplication 
rule (\ref{multrule}) is preserved, the index $n$ referring to the 
(odd) number of generators being multiplied.  Equation \ref{value3}
is similarly invariant, thus demonstrating $\CN$ contradictions.

To briefly summarize the main points:  Every GHZ state has
a special set of axes ({$z_k$}) that define an $N$-axis 
rotation group $SO(2)^{\otimes N}$ under which it transforms 
according to a 2D representation.  This means that the rotation 
of an individual qubit around its special axis transforms the 
$N$-particle state untraceably.  We have shown that this symmetry 
leads to an exponentially growing number ($\CN$, \Eq{viol}) of 
GHZ contradictions.  These contradictions involve detector settings
represented by operators in Fig. 1 that transform faithfully under
$SO(2)^{\otimes N}$.  Each compatible subset (N-S and E-W) obeys KS
operator identities representing all realizations of each 
contradiction, one for each joint eigenstate.  And each 
eigenstate transforms according to a distinct 2D representation.

\begin{thebibliography}{99}
\bibitem{binaries}  The binary labelling in \Eq{GHZ3} is redundant because
$\setnk$ and its complement label the same state.  Including ($\pm$) and 
excluding complements accounts for all $2^N$ states.
\bibitem{flip} More precisely, coordinate axes are rotated by $\pi$ 
about $x_k$, changing the directions of both $z_k$ and $y_k$.
\end{thebibliography}
\end{document}